\documentclass[12pt]{article}

\usepackage{physics}
\usepackage{graphicx}
\usepackage{amssymb}
\usepackage{amsthm}
\usepackage{amsmath}
\usepackage{dsfont}
\usepackage{bm}
\usepackage{mathrsfs}
\usepackage{bbold}
\usepackage{color}
\usepackage{hyperref}
\usepackage{times}

\def\nn{\nonumber} 
\newcommand{\be}{\begin{equation}}
\newcommand{\ee}{\end{equation}}
\newcommand{\bea}{\setlength\arraycolsep{2pt} \begin{eqnarray}}
\newcommand{\eea}{\end{eqnarray}}

\title{\large {\bf The trace of field equations for higher-derivative
gravity and an equality associating the Lagrangian density with
a divergence term}}
\date{}

\author{Jun-Jin Peng\footnote{corresponding author: pengjjph@163.com}
\, ,  Hua Li  \\\ \\
\small \sl School of Physics and Electronic Science,
\small \sl Guizhou Normal University,\\
\small Guiyang, Guizhou 550025, People's Republic of China \\
}

\begin{document}
\maketitle
\vspace{-5pt}

\begin{center}
\textbf{Abstract}
\end{center}
We figure out the explicit expression for the trace of the field equations
associated to generic higher derivative theories of gravity endowed with
Lagrangians depending upon the metric and its Riemann tensor, together with
arbitrary order covariant derivatives of the Riemann tensor. Then an
equality linking the Lagrangian density with the covariant divergence of
a vector field is put forward in terms of the trace of the field equations.
As a significant application, we particularly concentrate on a broad range of
higher derivative theories of gravity with the Lagrangian density constructed
from the contraction of the product for metric tensors with the product of
the Riemann tensors and the arbitrary order covariant derivatives of the
Riemann tensor. By utilizing the trace for the equations of motion, such a 
type of Lagrangian density is expressed as the covariant divergence of 
a vector field.


\voffset=-.90pt
\vspace{10pt}
\newpage

\section{Introduction}\label{one}

Higher derivative theories of gravity has garnered significant
attention in diverse respects, and a great deal of effort has been
put into understanding such theories
\cite{StelHD,StelGRG,Mye87,IGAv95,IyerWald,JKM94,XDee14,DyHi09,fRSF,
fRFT,NoOd11,Mode12,Camp14,BiTal15,BKM17,BCMM19}.
Particularly, an important research direction of higher derivative
theories of gravity is to construct multifarious solutions
with or without matter fields within the context of those theories
\cite{MigWil92,CaiOh06,HDBH15,LPPS15,KKZ17,BuCan17,CGMP22,AHMfR16}.
On the basis of the solutions, various significant applications such
as in cosmology, astrophysics, and black hole thermodynamics can be
further carried out
\cite{Mye88,NoOdi01,DeTe03,BMS06,CaiC07,NOOik17,ReSan19,QuPhD99}.
To seek the solutions, as well as to explore their properties relevant
to classical and quantum aspects of gravity, geometry and topology,
a primary task is to figure out the field equations. As is known,
within the framework of the Lagrangian-based theories of gravity, the
conventional approach to derive the equations of motion for gravitational
field is to vary the Lagrangian under consideration with respect to the
metric tensor. The gravitational field equations stemming from such a
treatment are usually expressed in terms of the functional derivative
of the Lagrangian with respect to the metric \cite{DyHi09,QuPhD99}.

Quite recently, within the work \cite{JJPEQM}, as an alternative
of the conventional approach, making use of the surface term
arising from the variation of the Lagrangian considered,
the author derived the equations
of motion for gravity theories that are invariant under diffeomorphisms.
Since the surface term plays an central role in the derivation for
the equations of motion and it constitutes the starting point of the
whole derivation, here the approach proposed in Ref. \cite{JJPEQM} is
referred to as the surface term based approach for convenience.
A prominent advantage for this approach is that resorting to the
surface term to derive the field equations provides a natural
means for ruling out the derivative of the Lagrangian density with
respect to the metric. Consequently, the
field equations are simplified greatly in contrast to the ones
obtained via the conventional approach. By making use of
the surface term based approach, the author presented the explicit
expressions for the equations of motion associated to general
diffeomorphism-invariant Lagrangians encompassing
the arbitrary order covariant derivatives of the Riemann tensor
in Ref. \cite{JJPEQM}.

On the basis of those equations of motion derived from the perspective
of the surface term, our primary goal in the present paper is to figure out
the explicit and compact expression for the trace of the field equations
associated to the generic matter field free theories of gravity admitting
diffeomorphism invariance symmetry. Then such an expression is utilized
to explore the structure of the Lagrangian density that is built from
the contraction of the product for metric tensors with the product of
the Riemann tensors and the arbitrary order covariant derivatives of the
Riemann tensor (see Eq. (\ref{Lspecas}) for this Lagrangian density).
In terms of the trace of the field equations, we observe that such a
type of Lagrangian density can be expressed as the covariant
divergence of a vector field like in the works \cite{OlR10,YX22}.

The present paper is organized as follows. Within Section \ref{two},
we obtain the explicit expression for the trace of the field
equations within the context of the diffeomorphism invariant
theories of gravity in the absence of matter fields. Then we put
forward an equality (\ref{LinDiv}) establishing the relation between
the Lagrangian density and the covariant divergence of a vector field
under the constraint (\ref{ConLinDiv}). Within Section
\ref{three}, to illustrate the results obtained in the previous
section, we consider an important application in the higher
derivative gravities endowed with the Lagrangian density constructed
from the contraction among the metric, the Riemann tensor and the
covariant derivatives of the Riemann tensor. We ultimately
summarize our main results in the last section.

\section{The trace for the field equations and the
Lagrangian density expressed in terms of the divergence
of a vector field}\label{two}

Within the present section, we will pay attention to the derivation
for the trace of the field equations associated to generic theories
of pure gravity accommodating diffeomorphism invariance. Based on
the explicit expression for the trace of the field equations, which
rules out the derivative of the Lagrangian density with respect to the
metric, we will delve into the association of the Lagrangian density
with the divergence of a vector field.

Without loss of
generality, we start with the diffeomorphism invariant
Lagrangian that is built out of the metric together with $i$th-order
$(i=0,1,\cdot\cdot\cdot,m)$ covariant derivatives
of the Riemann curvature tensor. It can be expressed as \cite{IyerWald}
\be
\mathcal{L}=\sqrt{-g}L\left(g^{\alpha\beta},
R_{\mu\nu\rho\sigma},\nabla_{\lambda_1}R_{\mu\nu\rho\sigma},
\cdot\cdot\cdot,\nabla_{\lambda_1}\cdot\cdot\cdot
\nabla_{\lambda_m}R_{\mu\nu\rho\sigma}\right)
\, , \label{LagCovR}
\ee
in which the inverse metric $g^{\alpha\beta}$ and the tensors
$\nabla_{\lambda_1}\cdot\cdot\cdot
\nabla_{\lambda_i}R_{\mu\nu\rho\sigma}$s involved in the functional
$L$ are treated as independent variables. It is worth mentioning
that here we release the symmetry restriction to all the indices
for the covariant
derivatives contained in the variable $\nabla_{\lambda_1}\cdot\cdot\cdot
\nabla_{\lambda_i}R_{\mu\nu\rho\sigma}$ unlike the situations considered in
\cite{IyerWald,JJPEQM}. By means of following the surface term based approach
proposed in Ref. \cite{JJPEQM} to derive the expression
${E}^{\mu\nu}$ for the equations of motion associated to the
Lagrangian (\ref{LagCovR}), one obtains the result without the necessity of
encompassing the derivative of the Lagrangian density with respect to
the metric, that is,
\be
{E}^{\mu\nu}={P}^{\mu\lambda\rho\sigma}
{R}^{\nu}_{~\lambda\rho\sigma}-2\nabla_{\rho}\nabla_{\sigma}
{P}^{\rho\mu\nu\sigma}
-\frac{1}{2}{L}g^{\mu\nu}+{E}^{\mu\nu}_{Z}
\, . \label{EoMhatL}
\ee
Within Eq. (\ref{EoMhatL}), the rank-four tensor ${P}^{\mu\nu\rho\sigma}$
is defined through
\be
{P}^{\mu\nu\rho\sigma}=\sum^{m}_{i=0}
(-1)^{i}
\nabla_{\lambda_{i}}\cdot\cdot\cdot\nabla_{\lambda_1}
Z_{(i)}^{\lambda_1\cdot\cdot\cdot\lambda_i\mu\nu\rho\sigma}
\, , \label{Pdef}
\ee
in which the rank-$(i+4)$ tensor
$Z_{(i)}^{\lambda_1\cdot\cdot\cdot\lambda_i\mu\nu\rho\sigma}$,
representing the derivative of the Lagrangian density with respect
to the variable $\nabla_{\lambda_1}\cdot\cdot\cdot
\nabla_{\lambda_i}R_{\mu\nu\rho\sigma}$, is given by
\be
Z_{(i)}^{\lambda_1\cdot\cdot\cdot\lambda_i\mu\nu\rho\sigma}
=Z_{(i)}^{\lambda_1\cdot\cdot\cdot\lambda_i\alpha\beta\gamma\lambda}
\Delta^{\mu\nu\rho\sigma}_{\alpha\beta\gamma\lambda}
=\frac{\partial{L}}{\partial
\nabla_{\lambda_1}\cdot\cdot\cdot
\nabla_{\lambda_i}R_{\mu\nu\rho\sigma}}
\, ,\label{Zidef}
\ee
with the $\Delta^{\mu\nu\rho\sigma}_{\alpha\beta\gamma\lambda}$ tensor
defined in terms of the Kronecker delta through
\be
\Delta^{\mu\nu\rho\sigma}_{\alpha\beta\gamma\lambda}
=\frac{1}{2}\left(\delta^{[\mu}_{\alpha}\delta^{\nu]}_{\beta}
\delta^{[\rho}_{\gamma}\delta^{\sigma]}_{\lambda}
+\delta^{[\rho}_{\alpha}\delta^{\sigma]}_{\beta}
\delta^{[\mu}_{\gamma}\delta^{\nu]}_{\lambda}\right)
\, . \label{Delttdef}
\ee
We point out that it is of great convenience to introduce the
$\Delta^{\mu\nu\rho\sigma}_{\alpha\beta\gamma\lambda}$ tensor
to exhibit the index symmetry as same as that of the Riemann tensor.
For instance, the index symmetry of the tensor
$Z_{(i)}^{\lambda_1\cdot\cdot\cdot\lambda_i\mu\nu\rho\sigma}$
gives rise to ${P}^{\mu\nu\rho\sigma}={P}^{\alpha\beta\gamma\lambda}
\Delta^{\mu\nu\rho\sigma}_{\alpha\beta\gamma\lambda}$, implying that
${P}^{\mu\nu\rho\sigma}={P}^{[\mu\nu][\rho\sigma]}
={P}^{[\rho\sigma][\mu\nu]}$. What is more, the second-rank
tensor ${E}^{\mu\nu}_{Z}$ in
Eq. (\ref{EoMhatL}), which is relative to the tensors
$Z_{(i)}^{\lambda_1\cdot\cdot\cdot\lambda_i\mu\nu\rho\sigma}$s
$(i=1,\cdot\cdot\cdot,m)$ and vanishes once the Lagrangian is
merely constructed out of the metric and its Riemann tensor,
takes the form
\be
{E}^{\mu\nu}_{Z}=\sum^m_{i=0}W^{\mu\nu}_{(i)}
\, , \label{EZdef}
\ee
where the second-rank tensor $W^{\mu\nu}_{(i)}$
$(i=0,1,\cdot\cdot\cdot,m)$ is defined through
\bea
W^{\mu\nu}_{(i)}&=&\frac{1}{2}
\sum^i_{k=1}(-1)^{k-1}\Big(\nabla_{\lambda_{k-1}}\cdot\cdot\cdot
\nabla_{\lambda_{1}}Z^{\lambda_1\cdot\cdot\cdot\lambda_{k-1}\mu
\lambda_{k+1}\cdot\cdot\cdot\lambda_{i}\alpha\beta\rho\sigma}_{(i)}
\Big) \nabla^\nu \nabla_{\lambda_{k+1}}
\cdot\cdot\cdot\nabla_{\lambda_{i}}R_{\alpha\beta\rho\sigma}
 \nn \\
&&+\frac{1}{2}\nabla_\lambda\left({U}^{(\mu\nu)\lambda}_{(i)}
-{U}^{\lambda(\mu\nu)}_{(i)}
+{U}^{[\mu|\lambda|\nu]}_{(i)}\right)
\, , \label{Widef}
\eea
with the third-rank tensor ${U}^{\mu\alpha\beta}_{(i)}$ defined in terms
of $Z_{(i)}^{\lambda_1\cdot\cdot\cdot\lambda_i\mu\nu\rho\sigma}$s as
\bea
{U}^{\mu\alpha\beta}_{(i)}&=&
4\sum^i_{k=1}(-1)^k\Big(\nabla_{\lambda_{k-1}}
\cdot\cdot\cdot\nabla_{\lambda_1}
Z_{(i)}^{\lambda_1\cdot\cdot\cdot
\lambda_{k-1}\alpha\lambda_{k+1}\cdot\cdot\cdot
\lambda_{i}\beta\gamma\rho\sigma}\Big)
\nabla_{\lambda_{k+1}}\cdot\cdot\cdot\nabla_{\lambda_i}
R^\mu_{~\gamma\rho\sigma}
\nn \\
&&+\sum^{i-1}_{k=1}\sum^i_{\ell=k+1}(-1)^k
\Big(\nabla_{\lambda_{k-1}}
\cdot\cdot\cdot\nabla_{\lambda_1}
Z_{(i)}^{\lambda_1\cdot\cdot\cdot\lambda_{k-1}
\alpha\lambda_{k+1}\cdot\cdot\cdot
\lambda_{\ell-1}\beta\lambda_{\ell+1}
\cdot\cdot\cdot\lambda_{i}
\tau\gamma\rho\sigma}\Big)\times \nn \\
&&\times \nabla_{\lambda_{k+1}}\cdot\cdot\cdot
\nabla_{\lambda_{\ell-1}}\nabla^\mu
\nabla_{\lambda_{\ell+1}}\cdot\cdot\cdot
\nabla_{\lambda_{i}}R_{\tau\gamma\rho\sigma}
\, . \label{Uidef}
\eea
Apparently, ${U}^{\mu\alpha\beta}_{(0)}=0$ and
${U}^{\mu\alpha\beta}_{(1)}
=-4{Z}^{\alpha\beta\gamma\rho\sigma}_{(1)}
{R}^\mu_{~\gamma\rho\sigma}$, and the former results in
$W^{\mu\nu}_{(0)}=0$. The anti-symmetric part of the tensor
${E}^{\mu\nu}_{Z}$ satisfies the identity
\be
{E}^{[\mu\nu]}_{Z}=-2{P}^{[\mu|\lambda\rho\sigma|}
{R}^{\nu]}_{~~\lambda\rho\sigma}
\, . \label{EZantip}
\ee
This gives rise to that ${E}^{\mu\nu}={E}^{\nu\mu}$. Here we point
out that the explicit expression (\ref{EoMhatL}) for equations of
motion is instrumental in our analysis throughout the present work.

On the basis of the expression (\ref{EoMhatL}) for the field equations,
implementing straightforward computations upon its trace, one arrives at
\be
{E}^{\mu}_{\mu}={P}^{\alpha\beta\rho\sigma}
{R}_{\alpha\beta\rho\sigma}-2{g}_{\rho\sigma}\nabla_{\mu}\nabla_{\nu}
{P}^{\mu\rho\sigma\nu}
-\frac{D}{2}{L}+{g}_{\mu\nu}{E}^{\mu\nu}_{Z}
\, . \label{TraEoML}
\ee
Here the positive integer $D$ stands for the dimensions of spacetimes.
Furthermore, by the aid of the following equality
\bea
\nabla_\mu\left(\sum^{n-1}_{\ell=0}
{A}^\mu_{(i,n;\ell)}\right)
&=&(-1)^{n}\Big(\nabla_{\lambda_{n}}\cdot\cdot\cdot
\nabla_{\lambda_{1}}Z^{\lambda_1\cdot\cdot\cdot
\lambda_{i}\alpha\beta\rho\sigma}_{(i)}\Big)\nabla_{\lambda_{n+1}}
\cdot\cdot\cdot\nabla_{\lambda_{i}}R_{\alpha\beta\rho\sigma}\nn\\
&&-Z^{\lambda_1\cdot\cdot\cdot\lambda_{i}\alpha\beta\rho\sigma}_{(i)}
\nabla_{\lambda_{1}}\cdot\cdot\cdot\nabla_{\lambda_{i}}
{R}_{\alpha\beta\rho\sigma}
\, , \label{IdentA}
\eea
where the vector ${A}^\mu_{(i,n;\ell)}$ $(n=1,2,\cdot\cdot\cdot,i;
\ell=0,1,\cdot\cdot\cdot,n-1)$ is defined as
\be
{A}^\mu_{(i,n;\ell)}=
(-1)^{n-\ell}\Big(\nabla_{\lambda_{n-\ell-1}}\cdot\cdot\cdot
\nabla_{\lambda_{1}}Z^{\lambda_1\cdot\cdot\cdot\lambda_{n-\ell-1}\mu
\lambda_{n-\ell+1}\cdot\cdot\cdot\lambda_{i}\alpha\beta\rho\sigma}_{(i)}
\Big)\nabla_{\lambda_{n-\ell+1}}
\cdot\cdot\cdot\nabla_{\lambda_{i}}R_{\alpha\beta\rho\sigma}
\, , \label{Ainelldef}
\ee
with ${A}^\mu_{(0,n;\ell)}\equiv0$ and
${A}^\mu_{(i,n;\ell)}={A}^\mu_{(i,p;q)}$ under the condition
$n-\ell=p-q$, the scalar ${P}^{\alpha\beta\rho\sigma}
{R}_{\alpha\beta\rho\sigma}$ can be put into the form
\bea
{P}^{\alpha\beta\rho\sigma}{R}_{\alpha\beta\rho\sigma}&=&
\sum^{m}_{i=0}
{Z}^{\lambda_1\cdot\cdot\cdot\lambda_{i}\alpha\beta\rho\sigma}_{(i)}
\nabla_{\lambda_{1}}\cdot\cdot\cdot\nabla_{\lambda_{i}}
{R}_{\alpha\beta\rho\sigma}
+\nabla_\mu\left(\sum^{m}_{i=0}\sum^{i-1}_{\ell=0}
{A}^\mu_{(i,i;\ell)}\right) \nn \\
&=&\sum^{m}_{i=0}
{Z}^{\lambda_1\cdot\cdot\cdot\lambda_{i}\alpha\beta\rho\sigma}_{(i)}
\nabla_{\lambda_{1}}\cdot\cdot\cdot\nabla_{\lambda_{i}}
{R}_{\alpha\beta\rho\sigma}
+\nabla_\mu\left(\sum^{m}_{i=0}\sum^{i}_{k=1}
{A}^\mu_{(i,i;i-k)}\right)
\, .\label{PRcontrac}
\eea
Apart from this, the scalar ${g}_{\mu\nu}{E}^{\mu\nu}_{Z}$ in
Eq. (\ref{TraEoML}) is able to be written as
\bea
{g}_{\mu\nu}{E}^{\mu\nu}_{Z}&=&
\frac{1}{2}\sum^{m}_{i=0}
i{Z}^{\lambda_1\cdot\cdot\cdot\lambda_{i}\alpha\beta\rho\sigma}_{(i)}
\nabla_{\lambda_{1}}\cdot\cdot\cdot\nabla_{\lambda_{i}}
{R}_{\alpha\beta\rho\sigma}
+\frac{1}{2}\nabla_\mu\left(\sum^{m}_{i=0}\sum^{i}_{k=1}
\sum^{k-2}_{\ell=0}
{A}^\mu_{(i,k-1;\ell)}\right) \nn \\
&&+\frac{1}{2}\nabla_\mu\left(
\sum^{m}_{i=0}{g}_{\rho\sigma}\Big({U}^{\rho\sigma\mu}_{(i)}
-{U}^{\mu\rho\sigma}_{(i)}\Big)\right)
\, . \label{TracEZ}
\eea
By plugging Eqs. (\ref{PRcontrac}) and (\ref{TracEZ}) back into
Eq. (\ref{TraEoML}), the trace relative to $E^{\mu\nu}$ is reformulated
as a compact form
\be
{E}^{\mu}_{\mu}=\frac{1}{2}\sum^{m}_{i=0}
(i+2){Z}^{\lambda_1\cdot\cdot\cdot\lambda_{i}\alpha\beta\rho\sigma}_{(i)}
\nabla_{\lambda_{1}}\cdot\cdot\cdot\nabla_{\lambda_{i}}
{R}_{\alpha\beta\rho\sigma}
-\frac{D}{2}{L}+\nabla_\mu{V}^\mu
\, , \label{TraEoML2}
\ee
with the vector ${V}^\mu$ given by
\bea
{V}^\mu&=&\frac{1}{2}\sum^{m}_{i=0}\left({Y}^\mu_{(i)}
+{g}_{\rho\sigma}{U}^{\rho\sigma\mu}_{(i)}
-{g}_{\rho\sigma}{U}^{\mu\rho\sigma}_{(i)}\right)
-2{g}_{\rho\sigma}\nabla_{\nu}{P}^{\mu\rho\sigma\nu}
\, . \label{Vdef} \quad
\eea
Within Eq. (\ref{Vdef}), the vector ${Y}^\mu_{(i)}$ has the form
\bea
{Y}^\mu_{(i)}&=&2\sum^{i}_{k=1}
{A}^\mu_{(i,i;i-k)}+\sum^{i}_{k=1}
\sum^{k-2}_{\ell=0}{A}^\mu_{(i,k-1;\ell)} \nn \\
&=&{g}_{\rho\sigma}{U}^{\rho\mu\sigma}_{(i)}
-2\sum^{i}_{k=1}{A}^\mu_{(i,k;0)}\nn \\
&=&\sum^{i}_{k=1}(i-k+2){A}^\mu_{(i,k;0)}
\, . \label{Yidef} 
\eea
For example, we have ${Y}^\mu_{(0)}=0$ and
${Y}^\mu_{(1)}=2{A}^\mu_{(1,1;0)}=-2Z_{(1)}^{\mu\alpha\beta\rho\sigma}
{R}_{\alpha\beta\rho\sigma}$. As a matter of fact, the vector field
${V}^\mu$ in Eq. (\ref{Vdef}) is non-unique and it is determined up
to an arbitrary divergence-free vector field. Particularly, let us
focus on a relatively simple case where the Lagrangian density $L$
is in absence of all the variables comprising the covariant
derivatives of the Riemann tensor, namely,
${L}=\bar{L}\big(g^{\alpha\beta},R_{\mu\nu\rho\sigma}\big)$.
In such a case, equation (\ref{EoMhatL}) returns to the expression for
equations of motion given by \cite{Pady}. Accordingly, the vector field
${V}^\mu$ reduces to the simple form
\be
\bar{V}^\mu=-2{g}_{\rho\sigma}\nabla_{\nu}
\frac{\partial\bar{L}}{\partial{R}_{\mu\rho\sigma\nu}}
\, . \label{Vmubar}
\ee

Equation (\ref{TraEoML2}) is our expected result. We stress that
the structure of this equation reveals that it is unnecessary to
deal with the second-rank tensor
$\frac{\partial{L}}{\partial{g}^{\mu\nu}}$ or
$\frac{\partial{L}}{\partial{g}_{\mu\nu}}$ resulting from the absence
of such a tensor in the field equations. As a side note,
here we point out that the authors also investigated the trace of the
field equations associated to the higher-derivative gravities
encompassing the covariant derivatives of the Riemann tensor
from a general perspective in Ref. \cite{OlR10}. By contrast,
firstly, unfortunately, the result of that work is incomplete
because of the incompleteness of the equations
of motion (see Eq. (A2) in \cite{OlR10}), which constitute the
foundation of their analysis and lack the contributions from the
variations of the covariant derivatives.\footnote{The complete field
equations for the Lagrangian (A1) in Ref. \cite{OlR10} are given
by Eq. (59) in Ref. \cite{JJPEQM}.} Secondly, the concrete
expression of the vector ${V}^\mu$ in the divergence term included
by the trace equation was absent within that work unlike our
explicit form. To tackle this problem, the complete field equations
have to be supplied. Thirdly, the field equations in that work contain the
derivative of the Lagrangian density with respect to the metric. The
existence of such a term renders it inconvenient to perform field equation
based analysis from general perspectives, although it does not affect
the trace of the field equations.

Specially, in light of Eq. (\ref{TraEoML2}), if the Lagrangian density
$L$ fulfills the constraint
\be
\sum^{m}_{i=0}
(i+2){Z}^{\lambda_1\cdot\cdot\cdot\lambda_{i}\alpha\beta\rho\sigma}_{(i)}
\nabla_{\lambda_{1}}\cdot\cdot\cdot\nabla_{\lambda_{i}}
{R}_{\alpha\beta\rho\sigma}=CL+2\nabla_\mu{B}^\mu
\, , \label{ConLinDiv}
\ee
where $C$ is some constant parameter and ${B}^\mu$ serves as a certain
vector field, the field equations ${E}^{\mu\nu}=0$ result in that the
Lagrangian density $L$ can be expressed as the covariant divergence of
a vector field in terms of the relation
\be
\frac{C-D}{2}L+\nabla_\mu\big({V}^\mu+{B}^\mu\big)=0
\, . \label{LinDiv}
\ee
As what has been shown in \cite{OlR10}, equation (\ref{LinDiv}) can
be adopted to classify the six-derivative Lagrangians of gravity.
Apart from this, it has been demonstrated in \cite{YX22,XWZ25} that
such an equation plays a crucial role in implementing Euclidean integrals
of gravitational actions \cite{HawPag} within the framework of
higher-derivative theories. Thanks to the significant applications, a typical
case of Eq. (\ref{LinDiv}) will be under consideration within the next
section. Interestingly,
when the dimension of spacetimes satisfies $D=C$, equation (\ref{LinDiv})
turns into $\nabla_\mu\big({V}^\mu+{B}^\mu\big)=0$, from which a conserved
vector field $J^\mu$ is extracted, written as
\be
J^\mu={V}^\mu+{B}^\mu
\, . \label{ConsVect}
\ee

\section{A significant application}\label{three}

Till now, higher derivative theories of gravity in the presence or
absence of matter fields have been continuing to be under intensive
scrutiny. Within the framework of a broad range of
such theories that are armed with diffeomorphism invariance symmetry
and free of matter fields, we take into consideration of
a significant application of Eq. (\ref{TraEoML2}).
In general, the Lagrangian densities of these theories usually allow
for the following scalar
\bea
\tilde{L}=\beta_{(k_0,k_1,\cdot\cdot\cdot,k_m)}
{g}^{\bullet\bullet}\cdot\cdot\cdot{g}^{\bullet\bullet}
\prod^m_{i=0}\left(\nabla_{\lambda_1}\cdot\cdot\cdot
\nabla_{\lambda_i}R_{\mu_i\nu_i\rho_i\sigma_i}\right)^{k_i}
\, . \label{Lspecas}
\eea
In the above equation, the arbitrary non-negative integer $k_i$
$(i=0,1,\cdot\cdot\cdot,m)$ is the number of the rank-$(i+4)$
tensor $\nabla_{\lambda_1}\cdot\cdot\cdot
\nabla_{\lambda_i}R_{\mu_i\nu_i\rho_i\sigma_i}$ included in the
Lagrangian density and $\beta_{(k_0,k_1,\cdot\cdot\cdot,k_m)}$
denotes any constant parameter. The second-rank tensor
${g}^{\bullet\bullet}$ represents the inverse metric tensor
$g^{\mu\nu}$, while the contravariant tensor
${g}^{\bullet\bullet}\cdot\cdot\cdot{g}^{\bullet\bullet}$,
representing the product of $\frac{1}{2}\sum^m_{i=0}(i+4)k_i$
inverse metric tensors, is required to possess the same indices
as the covariant one
$\prod^m_{i=0}\left(\nabla_{\lambda_1}\cdot\cdot\cdot
\nabla_{\lambda_i}R_{\mu_i\nu_i\rho_i\sigma_i}\right)^{k_i}$
thus to ensure that $\tilde{L}$ is a scalar field. This implies
that the production for all the inverse metric tensors is of
rank-$\sum^m_{i=0}(i+4)k_i$. More specifically,
the tensor $\left(\nabla_{\lambda_1}\cdot\cdot\cdot
\nabla_{\lambda_i}R_{\mu_i\nu_i\rho_i\sigma_i}\right)^{k_i}$,
standing for the multiplication of $k_i$ rank-$(i+4)$ tensors
comprised of the $i$th-order covariant derivatives of the Riemann
tensor, is written as
\bea
\left(\nabla_{\lambda_1}\cdot\cdot\cdot
\nabla_{\lambda_i}R_{\mu_i\nu_i\rho_i\sigma_i}\right)^{k_i}&=&
\prod^{k_i}_{p=1}\nabla_{\lambda_{(1,p)}}\cdot\cdot\cdot
\nabla_{\lambda_{(i,p)}}R_{\mu_{(i,p)}\nu_{(i,p)}\rho_{(i,p)}\sigma_{(i,p)}}
\, . \label{deliRki}
\eea
In particular, when $k_i=0$, it is implied that the
term $\left(\nabla_{\lambda_1}\cdot\cdot\cdot
\nabla_{\lambda_i}R_{\mu_i\nu_i\rho_i\sigma_i}\right)^{k_i}$ is absent
in the Lagrangian density. For the Lagrangian $\sqrt{-g}\tilde{L}$,
the rank-$(i+4)$ tensor
$\tilde{Z}^{\lambda_1\cdot\cdot\cdot\lambda_{i}\mu\nu\rho\sigma}_{(i)}
={Z}^{\lambda_1\cdot\cdot\cdot\lambda_{i}\mu\nu\rho\sigma}_{(i)}
\big|_{L\rightarrow\tilde{L}}=\frac{\partial{\tilde{L}}}{\partial
\nabla_{\lambda_1}\cdot\cdot\cdot\nabla_{\lambda_i}R_{\mu\nu\rho\sigma}}$
turns out to be the following explicit form
\bea
\tilde{Z}^{\lambda_1\cdot\cdot\cdot\lambda_{i}\mu\nu\rho\sigma}_{(i)}
&=&\beta_{(k_0,k_1,\cdot\cdot\cdot,k_m)}
{g}^{\bullet\bullet}\cdot\cdot\cdot{g}^{\bullet\bullet}
\left(\prod^m_{j=0, j\neq{i}}\big(\nabla_{\lambda_1}\cdot\cdot\cdot
\nabla_{\lambda_j}R_{\mu_j\nu_j\rho_j\sigma_j}\big)^{k_j} \right)\nn \\
&&\times\left(\sum^{k_i}_{q=1}
\delta^{\lambda_1}_{\lambda_{(1,q)}}\cdot\cdot\cdot
\delta^{\lambda_i}_{\lambda_{(i,q)}}
\Delta^{\mu\nu\rho\sigma}_{\mu_{(i,q)}\nu_{(i,q)}\rho_{(i,q)}\sigma_{(i,q)}}
\right. \nn \\
&&\left.
\times\prod^{k_i}_{p=1,p\neq{q}}\nabla_{\lambda_{(1,p)}}\cdot\cdot\cdot
\nabla_{\lambda_{(i,p)}}
{R}_{\mu_{(i,p)}\nu_{(i,p)}\rho_{(i,p)}\sigma_{(i,p)}}\right)
\, . \label{tildZdef}
\eea
By utilizing Eq. (\ref{tildZdef}), it can be proved that there exists
an equality connecting the Lagrangian density with its derivative
with respect to the covariant derivatives of the Riemann tensor,
being read off as
\be
k_i\tilde{L}=
\tilde{Z}^{\lambda_1\cdot\cdot\cdot\lambda_{i}\alpha\beta\rho\sigma}_{(i)}
\nabla_{\lambda_{1}}\cdot\cdot\cdot\nabla_{\lambda_{i}}
{R}_{\alpha\beta\rho\sigma}
\, . \label{tildLIdent}
\ee
As a consequence, the constant parameter $C$ in Eq. (\ref{ConLinDiv})
takes the value $\sum^{m}_{i=0}(i+2)k_i$,
and the divergence of the vector field ${B}^\mu$ vanishes. Then
substituting Eq. (\ref{tildLIdent}) into (\ref{LinDiv}) gives rise
to a much more compact formulation for the trace of the field equations
arising from the Lagrangian density (\ref{Lspecas}), taking the
following form
\be
\frac{1}{2}\left(\sum^{m}_{i=0}
(i+2)k_i-D\right)\tilde{L}+\nabla_\mu\tilde{V}^\mu=0
\, . \label{TraEoMtildL}
\ee
Within the above equation, the vector field $\tilde{V}^\mu$ is the
counterpart of the one ${V}^\mu$ in the situation for the Lagrangian
density (\ref{Lspecas}), that is,
$\tilde{V}^\mu={V}^\mu\big|_{Z\rightarrow\tilde{Z}}$, while the integer
number $\sum^{m}_{i=0}(i+2)k_i$, being the number of all the covariant 
derivatives plus the double of the number for all the Riemann tensors 
contained in $\tilde{L}$, has to be even, attributed to the fact
that the rank of the contravariant tensor
${g}^{\bullet\bullet}\cdot\cdot\cdot{g}^{\bullet\bullet}$ is even too.
Equation (\ref{TraEoMtildL}) demonstrates that the Lagrangian density
$\tilde{L}$ can be put into the covariant divergence of a vector
field in the on-shell case. Here we derive the complete form
for the equality (\ref{TraEoMtildL}) by means of explicit calculations
based on the field equations. Besides, in the paper \cite{OlR10},
the authors obtained this equality without the concrete expression
for the vector $\tilde{V}^\mu$ via computing the trace of the
field equations under the scaling analysis to the Lagrangian density,
given by Eq. (A9) in \cite{OlR10}. What is more, within the works
\cite{YX22,XWZ25}, under the decomposition for the variation of the
Lagrangian into three parts, corresponding to the determinant of the
metric, the inverse metric and the Riemann curvature tensor
(and/or the covariant derivative), respectively, the authors also
proposed equality (\ref{TraEoMtildL}) in absence of the concrete
expression for the vector $\tilde{V}^\mu$ like the situation in
\cite{OlR10}. In practice, the failure of those works in
giving the exact expression for $\tilde{V}^\mu$ originates from
the absence of the explicit field equations. We believe that the
complete expression for Eq. (\ref{TraEoMtildL}) could provide
great convenience for doing Euclidean integrals of gravitational
actions in diffeomorphism-invariant gravities as is shown
in \cite{YX22,XWZ25}.

To illustrate Eq. (\ref{TraEoMtildL}), we take into account a
Lagrangian $\sqrt{-g}{L}_1$ supposed to merely depend
on the metric and the Riemann tensor, together with the first-order
covariant derivative of the latter, where the Lagrangian density
${L}_1$ has the form
\bea
{L}_1&=&g^{\alpha_0\alpha}g^{\beta_0\beta}g^{\theta_0\theta}g^{\phi_0\phi}
g^{\kappa_0\kappa}g^{\gamma_0\gamma}g^{\lambda_0\lambda}
g^{\tau_0\tau}g^{\omega_0\omega}
{R}_{\alpha_0\beta_0\theta_0\phi_0}{R}_{\theta\phi\gamma\lambda}
\big(\nabla_{\kappa_0}{R}_{\gamma_0\lambda_0\tau_0\omega_0}\big)
\nabla_\kappa{R}_{\alpha\beta\tau\omega} \nn \\
&=&{R}^{\alpha\beta\theta\phi}{R}_{\theta\phi\gamma\lambda}
\big(\nabla_\kappa{R}_{\alpha\beta\tau\omega}\big)
\nabla^\kappa{R}^{\gamma\lambda\tau\omega}
\, . \label{Lden1def}
\eea
For the Lagrangian density (\ref{Lden1def}), both the tensors
${Z}^{\mu\nu\rho\sigma}_{(0)}$ and
${Z}^{\chi\mu\nu\rho\sigma}_{(1)}$ are given by
\bea
{Z}^{\mu\nu\rho\sigma}_{(1,0)}&=&
\frac{\partial{L}_1}{\partial{R}_{\mu\nu\rho\sigma}}=
2{\Delta}^{\mu\nu\rho\sigma}_{\theta\phi\gamma\lambda}
{R}^{\alpha\beta\theta\phi}
\big(\nabla_\kappa{R}_{\alpha\beta\tau\omega}\big)
\nabla^\kappa{R}^{\gamma\lambda\tau\omega} \, , \nn \\
{Z}^{\chi\mu\nu\rho\sigma}_{(1,1)}&=&
\frac{\partial{L}_1}{\partial\nabla_\chi{R}_{\mu\nu\rho\sigma}}=
2{\Delta}^{\mu\nu\rho\sigma}_{\alpha\beta\tau\omega}
{R}^{\alpha\beta\theta\phi}{R}_{\theta\phi\gamma\lambda}
\nabla^\chi{R}^{\gamma\lambda\tau\omega}
\, , \label{Z01forL1}
\eea
respectively. It is easy to confirm that
${Z}^{\mu\nu\rho\sigma}_{(1,0)}{R}_{\mu\nu\rho\sigma}=2L_1$
and ${Z}^{\chi\mu\nu\rho\sigma}_{(1,1)}\nabla_\chi
{R}_{\mu\nu\rho\sigma}=2L_1$. Moreover, substituting Eq. (\ref{Z01forL1})
into (\ref{Vdef}), one obtains the $V^\mu$ vector field associated to
the Lagrangian density (\ref{Lden1def}), presented by
\bea
{V}^\mu_1&=&-{Z}^{\mu\alpha\beta\rho\sigma}_{(1,1)}
{R}_{\alpha\beta\rho\sigma}-2{Z}^{\alpha\mu\beta\rho\sigma}_{(1,1)}
{R}_{\alpha\beta\rho\sigma}
+2{g}_{\rho\sigma}{Z}^{\rho\sigma\gamma\alpha\beta}_{(1,1)}
{R}^\mu_{~\gamma\alpha\beta} \nn \\
&&-2{g}_{\rho\sigma}\nabla_\alpha
{Z}^{\mu\rho\sigma\alpha}_{(1,0)}
+2{g}_{\rho\sigma}\nabla_\alpha\nabla_\beta
{Z}^{\beta\mu\rho\sigma\alpha}_{(1,1)}
\, . \label{Vmu1def}
\eea
As a consequence of Eq. (\ref{TraEoMtildL}), the Lagrangian density
${L}_1$ is related to the ${V}^\mu_1$ vector in the manner
\be
\frac{10-D}{2}{L}_1+\nabla_\mu{V}^\mu_1=0
\, . \label{TraEoMIDL1}
\ee
Particularly, in $D=10$ dimensions, the vector ${V}^\mu_1$ is conserved.
Besides, as another specific example to demonstrate the equality
(\ref{TraEoMtildL}), let us concentrate on its application in
the Lagrangian density $\alpha_5{L_5}$ under consideration
in \cite{XWZ25}, where $\alpha_5$ represents a constant parameter.
Here such a Lagrangian density is written as
\bea
{L}_2&=&\alpha_5{g}^{\mu\rho}{g}^{\nu\sigma}{g}^{\tau\omega}
{g}^{\kappa_0\kappa}{g}^{\alpha_0\alpha}{g}^{\beta_0\beta}
{g}^{\gamma_0\gamma}{g}^{\lambda_0\lambda}
{g}^{\theta\phi}{R}_{\mu\nu\rho\sigma}
{R}_{\tau\kappa_0\omega\alpha}{R}_{\alpha_0\beta_0\gamma_0\lambda_0}
\nabla_\theta\nabla_\phi{R}_{\kappa\beta\gamma\lambda} \nn \\
&=&\alpha_5{R}{R}^\kappa_\alpha{R}^{\alpha\beta\gamma\lambda}
\Box{R}_{\kappa\beta\gamma\lambda}
\, . \label{L2densit}
\eea
For such a Lagrangian density, $k_0=3$, $k_2=1$ and all the other
$k_i$s disappear. Hence equation (\ref{TraEoMtildL}) turns into
\be
\frac{10-D}{2}{L}_2+\nabla_\mu{V}^\mu_2=0
\, . \label{TraEoMIDL2}
\ee
That is to say, the trace equation identifies the Lagrangian density
${L}_2$ with a total divergence term through
${L}_2=\frac{2}{D-10}\nabla_\mu{V}^\mu_2$.
Within Eq. (\ref{TraEoMIDL2}), the vector field ${V}^\mu_2$ is expressed
as the following explicit form
\bea
{V}^\mu_2&=&{R}_{\alpha\beta\rho\sigma}
\nabla_\lambda{Z}^{\lambda\mu\alpha\beta\rho\sigma}_{(2,2)}
-\frac{3}{2}{Z}^{\mu\lambda\alpha\beta\rho\sigma}_{(2,2)}
\nabla_\lambda{R}_{\alpha\beta\rho\sigma}
+\frac{1}{2}g_{\rho\sigma}\big(
{U}^{\rho\sigma\mu}_{(2,2)}-{U}^{\mu\rho\sigma}_{(2,2)}\big)\nn \\
&&-2{g}_{\rho\sigma}\nabla_\alpha
{Z}^{\mu\rho\sigma\alpha}_{(2,0)}
-2{g}_{\rho\sigma}\nabla_\nu\nabla_\alpha\nabla_\beta
{Z}^{\beta\alpha\mu\rho\sigma\nu}_{(2,2)}
\, , \label{Vmu2def}
\eea
in which the third-rank tensor ${U}^{\mu\rho\sigma}_{(2,2)}$ is given by
\be
{U}^{\mu\rho\sigma}_{(2,2)}=
4R^\mu_{~\nu\alpha\beta}
\nabla_\lambda{Z}^{\lambda\rho\sigma\nu\alpha\beta}_{(2,2)}
-4{Z}^{\rho\lambda\sigma\nu\alpha\beta}_{(2,2)}
\nabla_\lambda R^\mu_{~\nu\alpha\beta}
-{Z}^{\rho\sigma\gamma\lambda\alpha\beta}_{(2,2)}
\nabla^\mu R_{\gamma\lambda\alpha\beta}
\, .  \label{U2def}
\ee
Within Eqs. (\ref{Vmu2def}) and (\ref{U2def}), after a
straightforward calculation, both the fourth-rank tensor
${Z}^{\mu\nu\rho\sigma}_{(2,0)}
=\frac{\partial{L}_2}{\partial{R}_{\mu\nu\rho\sigma}}$
and the sixth-rank one
${Z}^{\alpha\beta\mu\nu\rho\sigma}_{(2,2)}
=\frac{\partial{L}_2}{\partial\nabla_\alpha\nabla_\beta
{R}_{\mu\nu\rho\sigma}}$ are expressed respectively as
\bea
{Z}^{\mu\nu\rho\sigma}_{(2,0)}&=&
\alpha_5\Big(g^{\rho[\mu}g^{\nu]\sigma}{R}_{\alpha\kappa}
+{\Delta}^{\mu\nu\rho\sigma}_{\tau\alpha\omega\kappa}g^{\tau\omega}
{R}\Big){R}^{\alpha\beta\gamma\lambda}
\Box{R}^{\kappa}_{~\beta\gamma\lambda}
+\alpha_5{\Delta}^{\mu\nu\rho\sigma}_{\alpha\beta\gamma\lambda}
{R}{R}^\alpha_\kappa\Box{R}^{\kappa\beta\gamma\lambda} \, , \nn \\
{Z}^{\alpha\beta\mu\nu\rho\sigma}_{(2,2)}&=&\alpha_5
g^{\alpha\beta}{\Delta}^{\mu\nu\rho\sigma}_{\kappa\omega\gamma\lambda}
R{R}^{\kappa}_\tau{R}^{\tau\omega\gamma\lambda}
\, . \label{Z202def}
\eea
One is able to verify that
${Z}^{\mu\nu\rho\sigma}_{(2,0)}{R}_{\mu\nu\rho\sigma}=3L_2$
and ${Z}^{\alpha\beta\mu\nu\rho\sigma}_{(2,2)}\nabla_\alpha
\nabla_\beta{R}_{\mu\nu\rho\sigma}=L_2$. It is worth mentioning that the
equality (\ref{TraEoMIDL2}) was presented as well by means of the analysis
on the structure of the variation for the Lagrangian in the paper
\cite{XWZ25}. Nevertheless, the concrete expression for the vector
field ${V}^\mu_2$ was absent there owing to the complexity of such a
vector. What is more, the equality
(\ref{TraEoMIDL2}) can be reproduced by computing straightforwardly
the trace of the field equations for the Lagrangian with
$g^{\mu\nu}$, ${R}_{\mu\nu\rho\sigma}$ and $\Box{R}_{\mu\nu\rho\sigma}$
dependence given by \cite{PL2402}. In addition to this, performing
direct calculations on the trace of the field equations for the Lagrangian
$\sqrt{-g}{R}^{\mu\nu\rho\sigma}\Box^n{R}_{\mu\nu\rho\sigma}$ presented
by Eq. (165) in \cite{PL2402}, one can find that
equality (\ref{TraEoMtildL}) still holds.

Ultimately, let us move on to the linear combination for the Lagrangian
density ${L}_1$ with the one ${L}_2$, that is,
${L}_{[1,2]}=\lambda{L}_1+\chi{L}_2$, where $\lambda$ and
$\chi$ stand for two arbitrary constant parameters. From
$2\frac{\partial{L}_{[1,2]}}{\partial{R}_{\mu\nu\rho\sigma}}
{R}_{\mu\nu\rho\sigma}
+3\frac{\partial{L}_{[1,2]}}{\partial\nabla_\gamma{R}_{\mu\nu\rho\sigma}}
\nabla_\gamma{R}_{\mu\nu\rho\sigma}
+4\frac{\partial{L}_{[1,2]}}{\partial\nabla_\gamma\nabla_\lambda
{R}_{\mu\nu\rho\sigma}}\nabla_\gamma\nabla_\lambda{R}_{\mu\nu\rho\sigma}
=10{L}_{[1,2]}$, we observe that the fact that each of ${L}_1$ and
${L}_2$ satisfies the condition (\ref{ConLinDiv}) accordingly implies
that their linear combination ${L}_{[1,2]}$ satisfies such a condition 
as well. However, this does not always holds true. As a concrete 
counterexample, we take into consideration of the well-known Starobinsky 
model described by the Lagrangian density ${f}=\lambda{R}+\chi{R}^2$, 
which can be also interpreted as a special case of $f(R)$ gravity or 
quadratic curvature gravity. Although each of the Ricci scalar ${R}$ 
and its square ${R}^2$ separately obeys the condition (\ref{ConLinDiv}), 
according to $2\frac{\partial{f}}{\partial{R}_{\mu\nu\rho\sigma}}
{R}_{\mu\nu\rho\sigma}=4{f}-2\lambda{R}$ or
$2\frac{\partial{f}}{\partial{R}_{\mu\nu\rho\sigma}}
{R}_{\mu\nu\rho\sigma}=2{f}+2\chi{R}^2$, the existence of the additional
term $-2\lambda{R}$ or $2\chi{R}^2$ makes the Lagrangian density
${f}$ fail to fulfill the condition (\ref{ConLinDiv}) for arbitrary
$\lambda$ and $\chi$, attributed to the fact that the Ricci scalar
$R$ can not be associated to $R^2$ in the way that 
${R}=\kappa{R}^2+\nabla_\mu(\bullet)^\mu$ or
${R}^2=\kappa{R}+\nabla_\mu(\bullet)^\mu$ for arbitrary spacetime 
metric. Here and in what follows, $\kappa$ and $(\bullet)^\mu$ are 
adopted to represent some constant parameter and a certain vector 
field, respectively. Furthermore, for the purpose of generality,
we focus on the linear combination for the scalars $\tilde{L}$ and
$\tilde{L}'=\tilde{L}(\beta\rightarrow\beta', {k}\rightarrow{k}')$,
leading to
\be
\check{L}=\lambda\tilde{L}+\chi\tilde{L}'
\, . \label{LCofLLprim}
\ee
Introducing two integers $N=\sum^{m}_{i=0}(i+2)k_i$ and
$N'=\sum^{m}_{i=0}(i+2)k'_i$ to represent respectively the number of 
all the covariant derivatives plus the double of the number for all 
the Riemann tensors contained within $\tilde{L}$ and $\tilde{L}'$, 
we obtain
\bea
\sum^{m}_{i=0}
(i+2)\frac{\partial{\check{L}}}{\partial
\nabla_{\lambda_1}\cdot\cdot\cdot\nabla_{\lambda_i}R_{\mu\nu\rho\sigma}}
\nabla_{\lambda_{1}}\cdot\cdot\cdot\nabla_{\lambda_{i}}
{R}_{\mu\nu\rho\sigma}&=&N'\check{L}+(N-N')\lambda\tilde{L} \nn \\
&=&N\check{L}+(N'-N)\chi\tilde{L}'
\, . \label{PacheckL}
\eea
Apparently, under the condition that $N=N'$, equation (\ref{PacheckL}) 
completely fulfills the condition (\ref{ConLinDiv}). This further 
guarantees that equation (\ref{LinDiv}) arising from the trace of
the field equations holds for the Lagrangian density $\check{L}$. 
Actually, the aforementioned linear combination for ${L}_1$ and 
${L}_2$ belongs to such a situation. On the other hand, under the 
condition that $N\neq N'$, when neither of the scalars $\tilde{L}$ and 
$\tilde{L}'$ can be put into the form
$\kappa\check{L}+\nabla_\mu(\bullet)^\mu$, which in turn implies that 
both $\tilde{L}$ and $\tilde{L}'$ can not be related to each other through
$\tilde{L}=\kappa\tilde{L}'+\nabla_\mu(\bullet)^\mu$ or 
$\tilde{L}'=\kappa\tilde{L}+\nabla_\mu(\bullet)^\mu$, 
equation (\ref{PacheckL}) violates the condition (\ref{ConLinDiv}). 
Consequently, equation (\ref{LinDiv}) specific to $\check{L}$ breaks 
down. In such a situation, this equation is able to be modified as
\be
\frac{N-D}{2}\check{L}
+\nabla_\mu\big({V}^\mu|_{{L}\rightarrow\check{L}}\big)=
\frac{N-N'}{2}\chi\tilde{L}'
\,  \label{LcheckinDiv}
\ee
by adding a non-zero term to the right hand side of the equation.
By the way, here we point out that the Lagrangian density 
$f=\lambda{R}+\chi{R}^2$ belongs to the latter case where $N\neq N'$. 
Besides, for all the six-derivative Lagrangians of gravity constructed
in the paper \cite{OlR10}, all the Lagrangian densities are the linear 
combination of some scalars, each of which can be put into the form 
(\ref{Lspecas}) and takes the same $N=6$. Thus, unlike the Starobinsky 
model, each of these Lagrangian densities obeys equations (\ref{ConLinDiv})
and (\ref{LinDiv}) as is shown there.

\section{Summary}\label{four}

In closing, let us summarize the main results. Within the framework of
the higher derivative theories of gravity described by the general
diffeomorphism invariant Lagrangian (\ref{LagCovR}) built from the metric
and its Riemann tensor together with the covariant derivatives of the
Riemann curvature tensor, beginning with the expression for the equations
of motion, we obtain its trace presented by equation (\ref{TraEoML2}).
On the basis of equation (\ref{TraEoML2}), it is demonstrated that
there exists the identity (\ref{LinDiv}) connecting the Lagrangian
density with a divergence term under the sufficient condition
(\ref{ConLinDiv}). Our particular focus is on the significant
application of such an equality in the higher derivative theories of
gravity composed of the Lagrangian density (\ref{Lspecas}).
We further derive the trace equation (\ref{TraEoMtildL}) associated to
this Lagrangian density. The equality (\ref{TraEoMtildL}) renders it
feasible to express the Lagrangian density (\ref{Lspecas}) as the
divergence of a vector field.

Apparently, our analysis in the present work only involves theories
of gravity in the absence of matter fields, and at the heart of our
analysis is the expression (\ref{EoMhatL}) for the equations of
motion. For the purpose of universality, we observe that the
worthwhile future task is to generalize equations (\ref{TraEoML2})
and (\ref{LinDiv}) to the higher derivative theories of gravity in
the presence of matter fields. The achievement of such a task
strongly depends upon the generalization of equation (\ref{EoMhatL})
in those theories. Apart from this, equations (\ref{TraEoML2}) and
(\ref{TraEoMtildL}) may be applied to construct the static
spherically-symmetric solutions in generic higher derivative
theories of gravity admitting dffeomorphism invariance. What is more,
as another interesting extension, equality (\ref{TraEoMtildL}) can
be utilized to investigate the Euclidean integrals of gravitational
actions in the context of a wide range of higher derivative theories
of gravity like in \cite{YX22,XWZ25}. We expect to return to these
and relevant issues in future work.

\section*{Acknowledgments}

This work was supported by the National Natural Science
Foundation of China under Grant Nos. 11865006 and 12565009.

\end{document}